\documentclass[10pt,aps,twocolumn,preprintnumbers,prd,noshowpacs,nofootinbib,noshowkeys,floatfix,superscriptaddress,raggedbottom]{revtex4}
\usepackage[dvips]{graphics,graphicx}
\usepackage[colorlinks=true,linktocpage=true,linkcolor=blue,citecolor=blue]{hyperref}
\usepackage[usenames,dvipsnames]{color}
\usepackage{amsmath, amssymb,oldgerm}
\usepackage{multirow}
\usepackage{longtable}
\usepackage{color}
\usepackage[normalem]{ulem}  
\usepackage{braket}
\usepackage{slashed}
\usepackage[mathscr]{eucal}
\usepackage{amsmath}
\newcommand{\dif}{\mathrm{d}}
\newcommand{\vecs}[1]{\mathbf{#1}} 

\newcommand{\PB}[1]{\left\{#1\right\}_{\text{PB}}}
\newcommand{\DB}[1]{\left\{#1\right\}_{\text{DB}}}
\newcommand{\GPB}[1]{\left\{#1\right\}_{\text{GPB}}}

\let\originalsout\sout
\renewcommand{\sout}[1]{\textcolor{red}{\originalsout{#1}}}

\newcommand{\prlheading}[1]{\par\addvspace{5pt}\textit{#1.}---\ignorespaces}

\makeatletter
\AtBeginDocument{%
    \g@addto@macro\normalsize{%
        \setlength{\abovedisplayskip}{8pt plus 1pt minus 1pt}%
        \setlength{\belowdisplayskip}{8pt plus 1pt minus 1pt}%
        \setlength{\abovedisplayshortskip}{8pt plus 1pt minus 1pt}%
        \setlength{\belowdisplayshortskip}{8pt plus 1pt minus 1pt}%
    }%
    \normalsize
}
\makeatother

\begin{document}
\makeatletter
\@booleantrue\raggedcolumn@sw
\makeatother
\raggedbottom

\title{Constrained Hamiltonian Formulation of Kinetic Theory
\\ for Relativistic Spin-$1/2$ Particles}
\author{Jie Zhou}
\affiliation{State Key Laboratory of Dark Matter Physics, Shanghai Key Laboratory for Particle Physics and Cosmology,
	Key Laboratory for Particle Astrophysics and Cosmology (MOE),
	School of Physics and Astronomy, Shanghai Jiao Tong University, Shanghai 200240, China}
\author{Ying Shan Zhao}
\affiliation{State Key Laboratory of Dark Matter Physics, Shanghai Key Laboratory for Particle Physics and Cosmology,
	Key Laboratory for Particle Astrophysics and Cosmology (MOE),
	School of Physics and Astronomy, Shanghai Jiao Tong University, Shanghai 200240, China}
\author{Yifeng Sun}
\email[Contact author: ]{sunyfphy@sjtu.edu.cn}
\affiliation{State Key Laboratory of Dark Matter Physics, Shanghai Key Laboratory for Particle Physics and Cosmology,
	Key Laboratory for Particle Astrophysics and Cosmology (MOE),
	School of Physics and Astronomy, Shanghai Jiao Tong University, Shanghai 200240, China}

\date{\today}

\begin{abstract}
	Describing spin transport away from equilibrium requires a kinetic theory that consistently incorporates spin dynamics and phase space structure. We construct such a collisionless kinetic theory for relativistic spin-$1/2$ particles in external electromagnetic fields using the constrained Hamiltonian formalism. The formulation yields both a covariant kinetic equation and its form in independent physical phase space variables, together with the associated noncanonical Poisson brackets and invariant measure. For massive particles, the interplay between the spin constraints and the electromagnetic field makes the particle velocity generally nonparallel to its kinetic momentum and modifies the momentum and spin evolution. The invariant measure in the reduced phase space obeys an exact Liouville identity even for time-dependent fields. This identity and the equations of motion give local conservation of charge and Boltzmann entropy, as well as balance equations for energy--momentum and total angular momentum. The global equilibrium conditions are consistent with covariant statistical mechanics and the Wigner function approach. The massless construction recovers the standard chiral kinetic equations and phase space measure. This framework provides a consistent Hamiltonian description of spin transport beyond equilibrium.
\end{abstract}

\maketitle

\prlheading{Introduction}
Spin manifests itself in a rich variety of intriguing phenomena, such as the chiral magnetic effect, spin Hall effects, and spin polarization and alignment, across diverse physical systems \cite{Dyakonov:1971,Kato:2004,Wunderlich:2005,Marklund:2007,Hosten:2008,Kharzeev:2007jp,Fukushima:2008xe,Vlasenko:2014,STAR:2017ckg,STAR:2023SpinAlignment}. Understanding these phenomena beyond equilibrium requires a kinetic description that incorporates spin dynamics. Such a description is relevant to a wide range of physical systems, including laser-excited magnetic materials \cite{Beaurepaire:1996}, heavy-ion collisions \cite{Kurkela:2015,Hidaka:2024SpinRelaxation}, and the neutrino-rich environments of core-collapse supernovae and neutron star mergers \cite{Vlasenko:2013fja,Yamamoto:2023okm}.

Several approaches have been developed to describe spin kinetic theory of relativistic spin-$1/2$ fermions, with particular attention to the effects of external electromagnetic fields, including Wigner function methods based on gradient expansions \cite{Weickgenannt:2019dks,Liu_2020,Hidaka:2022dmn}, semiclassical actions incorporating Berry phases for chiral and massive particles \cite{Stephanov:2012ki,Son:2012wh,Chen:2013iga}, symplectic formulations of massless spinning particles \cite{Duval:2014ppa}, Foldy--Wouthuysen transformations of the Dirac Hamiltonian \cite{Ekman:2017kxi}, worldline formulations with Grassmann spin variables \cite{Mueller:2019gjj}, covariant Berry connections \cite{Stone:2014fja}, and bottom-up formulations based on spin-dependent Poisson brackets \cite{Mo:2025nes}. Reference~\cite{DERIGLAZOV20141} describes relativistic massive spin-$1/2$ particles in external electromagnetic fields within the constrained Hamiltonian formalism, yielding covariant equations of motion consistent with the physical spin constraints.

In this Letter, we show that the spin constraints also induce a noncanonical Poisson structure and modify the invariant phase space measure, providing the basis for a consistent collisionless kinetic description. Starting from a Lorentz scalar Hamiltonian with magnetic dipole coupling neglected, we formulate the dynamics in an extended canonical phase space of spacetime coordinates, conjugate momenta, and auxiliary spin variables. Imposing the mass shell and spin constraints yields covariant equations of motion, from which we construct the corresponding collisionless kinetic equation. Gauge fixing then gives its reduced form in terms of independent physical phase space variables, with the Poisson structure and measure determined by their Dirac brackets. The massless construction recovers the standard chiral kinetic equations and phase space measure. For massive particles, we establish an exact Liouville identity that ensures local charge and Boltzmann entropy conservation within the collisionless dynamics. Together with the particle equations of motion, this identity also yields the energy-momentum and total angular momentum balance equations. Our global equilibrium analysis yields conditions consistent with covariant statistical mechanics \cite{Becattini:2012StatMech,Florkowski:2018} and the Wigner-function approach \cite{Weickgenannt:2019dks,Liu_2020}.

\par Our notation and conventions are: $\hbar=c=1$, $a\cdot b=a^\mu b_\mu$, $a_{[\mu}b_{\nu]}=a_\mu b_\nu-a_\nu b_\mu$, $a_{(\mu}b_{\nu)}=a_\mu b_\nu+a_\nu b_\mu$, $g_{\mu\nu}=\mathrm{diag}(+,-,-,-)$, and $\epsilon^{0123}=-\epsilon_{0123}=1$. Repeated spatial indices are summed with the Euclidean metric, and the three-dimensional Levi-Civita symbol is defined with $\epsilon_{123}=1$.

\prlheading{Hamiltonian formulation of spin kinetic theory}
Following Refs. \cite{DERIGLAZOV2012309,DERIGLAZOV20141}, we describe each particle by a trajectory $\zeta^a(\lambda)=(x^\mu,p^\mu,\omega^\mu,\pi^\mu)$ in an extended canonical phase space, parametrized by a worldline parameter $\lambda$. Here $(x^\mu,p^\mu)$ describe spacetime coordinates and conjugate momenta, while $(\omega^\mu,\pi^\mu)$ are auxiliary canonical variables representing spin. Their nonvanishing fundamental Poisson brackets are
\begin{equation}
	\{x^\mu,p^\nu\}_{\mathrm{PB}}=g^{\mu\nu},
	\qquad
	\{\omega^\mu,\pi^\nu\}_{\mathrm{PB}}=g^{\mu\nu}.
	\label{eq:fundamental-pb}
\end{equation}
The spin tensor is expressed as
\begin{equation}
	S^{\mu\nu}=2\omega^{[\mu}\pi^{\nu]}.
	\label{eq:spin-tensor}
\end{equation}
The physical spin angular momentum tensor is $S^{\mu\nu}/2$; its Poisson brackets realize the classical Poisson relations of the Lorentz algebra.

The dynamics is governed by a Lorentz scalar Hamiltonian constructed solely from constraint functions,
\begin{equation}
	\mathcal H=\lambda_iT_i,
	\label{eq:constraint-hamiltonian}
\end{equation}
\noindent where $\lambda_i$ are Lagrange multipliers and $T_i\approx0$ implement the physical mass shell and spin constraints in canonical variables. For massive particles without a magnetic moment, the physical constraints are
\begin{equation}
	P^2=m^2,
	\qquad
	P_\mu S^{\mu\nu}=0,
	\qquad
	S^{\mu\nu}S_{\mu\nu}=8s^2,
	\label{eq:physical-constraints}
\end{equation}
\noindent where $P^\mu=p^\mu-eA^\mu$ is the kinetic momentum, $e$ is the particle charge, and $A^\mu$ is the external electromagnetic gauge potential. The second relation is the Tulczyjew--Dixon spin supplementary condition \cite{Tulczyjew1959,Dixon1974}, and the third specifies a fixed spin magnitude. For massive spin-$1/2$ particles, one can choose $s^2=3/4$ to match the quantum spin Casimir \cite{DERIGLAZOV20141,Weickgenannt:2020aaf}. The spin constraints~\eqref{eq:physical-constraints} leave two independent spin variables.

A magnetic dipole coupling would modify the mass shell condition $P^2-\mu eF^{\mu\nu}S_{\mu\nu}/2-m^2=0$, where $\mu=g/2$ and $g$ is the Landé $g$ factor. We focus on $\mu=0$, for which the interplay between the spin constraints and the external electromagnetic field already produces nontrivial effects. The massless case additionally requires spin constraints selecting a fixed helicity, rather than simply setting $m=0$, as discussed in the section on chiral kinetic theory.

The Dirac--Bergmann algorithm enforces constraint preservation, derives any further constraints from the consistency conditions, and separates the complete set into first and second class sets \cite{Wipf,Brown:2022gha}. Eliminating the second class constraints gives the partially reduced Hamiltonian $\mathcal H_{\mathrm{pr}}$ and the corresponding Dirac bracket, so that $\dif F/\dif\lambda=\{F,\mathcal H_{\mathrm{pr}}\}_{\mathrm{DB}}$ for any phase space function $F$ without explicit dependence on $\lambda$. The remaining first class constraints leave gauge freedom in the equations of motion. The resulting particle trajectories determine the covariant collisionless evolution of the phase space distribution function,
\begin{equation}
	\frac{\partial f}{\partial\lambda}
	+\frac{\dif\zeta^a}{\dif\lambda}
	\frac{\partial f}{\partial\zeta^a}=0.
	\label{eq:covariant-kinetic}
\end{equation} 
After gauge fixing, the full Dirac brackets define the generalized Poisson brackets (GPBs) of the independent variables on the reduced phase space,
$\Pi^{ab}=\{\zeta_r^a,\zeta_r^b\}_{\mathrm{GPB}}
=\{\zeta_r^a,\zeta_r^b\}_{\mathrm{DB,full}}$,
and determine the associated phase space measure. The Dirac bracket definition and the detailed constraint analysis are given in the Supplemental Material.

\prlheading{Chiral kinetic theory}
We now reconstruct chiral kinetic theory and its invariant phase space measure within this framework. For massless fermions, the conditions $P^2=0$, $P_\mu S^{\mu\nu}=0$ and $S^{\mu\nu}S_{\mu\nu}=8s^2$ are supplemented by \cite{ChenSonStephanov2015},
\begin{equation}
	N_\mu S^{\mu\nu}=0,
	\label{eq:spin-frame}
\end{equation}
\noindent where $N^\mu$ is a constant future-directed unit timelike auxiliary vector. These conditions determine the spin tensor as
\begin{equation}
	S^{\mu\nu}
	=\lambda_h\frac{\epsilon^{\mu\nu\rho\sigma}P_\rho N_\sigma}{P\cdot N},
	\label{eq:massless-spin}
\end{equation}
\noindent where $\lambda_h$ labels the two helicity states. Thus, no independent continuous spin variables remain, and no separate equation for spin evolution is required, since it is fully determined by the momentum dynamics at fixed helicity. Choosing $N^\mu=(1,0,0,0)$ recovers the usual helicity representation.

In the extended canonical phase space, the eight constraints are
\begin{align}
			&\{T_1,\ldots,T_8\}\notag\\
			=&\left\{
			\frac12P^2,\,
			\frac12(\pi^2-a_2),\,
			\frac12(\omega^2-a_3),\,P\cdot\omega,\,P\cdot\pi,\,\omega\cdot\pi,\right.\notag\\
			&\left.N\cdot\omega,\,N\cdot\pi\right\},
					\label{eq:massless-constraints}
	\end{align}
\noindent where $a_2$ and $a_3$ fix the normalization $S^{\mu\nu}S_{\mu\nu}=8a_2a_3$. The Dirac--Bergmann analysis identifies six second class constraints and two first class constraints, as detailed in the Supplemental Material (\ref{SM:ckt}). 

After imposing the second class constraints and redefining the remaining multipliers, the Hamiltonian reduces to
\begin{equation}
	\mathcal H_{\mathrm{pr}}
	=\frac{\lambda_1}{2}P^2
	+\frac{\lambda_2}{2}(\pi^2-a_2),
	\label{eq:massless-hamiltonian}
\end{equation}
\noindent where $\lambda_1$ and $\lambda_2$ are the remaining undetermined multipliers associated with the gauge freedoms of worldline reparametrization and the auxiliary spin variables, respectively. The resulting covariant equations of motion are
\begin{subequations}\label{eq:massless-covariant}
	\begin{gather}
		\frac{\dif x^\mu}{\dif \lambda}
		=\lambda_1\left(
		P^\mu-\frac{e}{2\Delta}F_{\rho\sigma}P^\sigma S^{\mu\rho}
		\right),
		\label{eq:massless-cov-x}
		\\
		\frac{\dif P^\mu}{\dif \lambda}
		=eF^{\mu\nu}\frac{\dif x_\nu}{\dif \lambda},
		\label{eq:massless-cov-p}
	\end{gather}
\end{subequations}
\noindent where $\Delta=(P\cdot N)^2-eN^2F_{\mu\nu}S^{\mu\nu}/4$.  These equations depend only on $\lambda_1$. The freedom associated with $\lambda_2$ arises from the redundant representation of physical spin by $(\omega^\mu,\pi^\mu)$, thus affecting neither physical evolution nor the generalized Poisson brackets between physical variables.

Choosing coordinate time as the worldline parameter $x^0=\lambda=t$ (so that $\dif x^0/\dif\lambda=1$), and taking $N^\mu=(1,0,0,0)$, the covariant equations reproduce the standard chiral kinetic equations of motion \cite{Stephanov:2012ki},
\begin{subequations}\label{eq:ckt-eom}
	\begin{align}
		\sqrt G\,\dot{\mathbf x}
		&=\hat{\mathbf P}
		+e\,\mathbf E\times\boldsymbol\Omega
		+e\,(\hat{\mathbf P}\cdot\boldsymbol\Omega)\,\mathbf B,\label{eq:ckt-x}\\
		\sqrt G\,\dot{\mathbf P}
		&=e\,\mathbf E
		+e\,\hat{\mathbf P}\times\mathbf B
		+e^2(\mathbf E\cdot\mathbf B)\boldsymbol\Omega,
		\label{eq:ckt-p}
	\end{align}
\end{subequations}
\noindent where dots now denote derivatives with respect to $t$, and
\begin{equation}
	\hat{\mathbf P}=\frac{\mathbf P}{|\mathbf P|},
	\qquad
	\boldsymbol\Omega
	=\frac{\lambda_h\mathbf P}{2|\mathbf P|^3},
	\qquad
	\sqrt G=1+e\,\mathbf B\cdot\boldsymbol\Omega.
	\label{eq:ckt-definitions}
\end{equation}
After fixing the remaining gauge freedom of the auxiliary spin variables by $\omega^1=0$, the generalized Poisson brackets among the physical reduced phase space variables $(\mathbf x,\mathbf P)$ coincide with those of chiral kinetic theory \cite{Duval:2005vn,Son:2012zy} (see Supplemental Material \ref{SM:ckt}) and directly determine the invariant measure on the reduced phase space  $\dif\Gamma=|\sqrt G|\,\dif^3\mathbf x\,\dif^3\mathbf P$. These brackets are independent of the auxiliary spin gauge choice, provided the first and second class constraints, together with the gauge conditions, form a complete second class set. Thus, the constrained Hamiltonian formulation consistently recovers both the particle dynamics and the phase space measure.

\prlheading{Spin kinetic theory for massive spin-1/2 particles}
For massive spin-1/2 fermions, the six constraints in the extended canonical phase space are
\begin{align}
	&\{T_1,\ldots,T_6\} \notag\\
	=&\left\{\dfrac12(P^2-m^2),\,
	\dfrac12(\pi^2-a_2),\,
	\dfrac12(\omega^2-a_3),\,P\cdot\omega,\,P\cdot\pi,\right. \notag\\
	&\left.\omega\cdot\pi\right\},
	\label{eq:massive-constraints}
\end{align}
with the same spin normalization $S^{\mu\nu}S_{\mu\nu}=8a_2a_3$. The Dirac--Bergmann analysis yields four second class constraints and two first class constraints (see Supplemental Material, Sec. \ref{sm:skm}). 

After imposing the second class constraints and redefining the remaining multipliers, the Hamiltonian reduces to
\begin{equation}
	\mathcal H_{\mathrm{pr}}
	=\frac{\lambda_1}{2}(P^2-m^2)
	+\frac{\lambda_2}{2}(\pi^2-a_2).
	\label{eq:massive-hamiltonian}
\end{equation}
The multipliers correspond to the same gauge freedoms as in the massless case. Although the evolution of the auxiliary spin variables depends on $\lambda_2$, the physical equations do not. Introducing the spin four-vector
\begin{equation}
	S^\mu
	=\frac{1}{2\sqrt{P^2}}
	\epsilon^{\mu\nu\alpha\beta}P_\nu S_{\alpha\beta},
	\label{eq:spin-vector}
\end{equation}
the covariant equations of motion become
\begin{subequations}\label{eq:massive-covariant}
	\begin{gather}
		\frac{\dif x^\mu}{\dif \lambda}
		=\lambda_1\left(
		P^\mu
		+\frac{e}{2\Delta_1\sqrt{P^2}}
		\epsilon^{\mu\nu\alpha\beta}P_\alpha S_\beta
		F_{\nu\rho}P^\rho
		\right),
		\label{eq:massive-cov-x}
		\\
		\frac{\dif P^\mu}{\dif\lambda}
		=eF^{\mu\nu}\frac{\dif x_\nu}{\dif \lambda},
		\label{eq:massive-cov-p}
		\\
		\frac{\dif S^\mu}{\dif\lambda}
		=-\frac{P^\mu}{P^2}S^\nu\frac{\dif P_\nu}{\dif \lambda},
		\label{eq:massive-cov-spin}
	\end{gather}
\end{subequations}
\noindent where $\Delta_1=m^2+eF_{\mu\nu}S^{\mu\nu}/4$.

We again choose coordinate time as the worldline parameter, $x^0=\lambda=t$. To avoid confusion with Minkowski index lowering, spatial components are denoted below by lower indices, with $S_i\equiv S^i$ and $P_i\equiv P^i$ for $i=1,2,3$. Taking $\mathbf S_r=(S_2,S_3)$ as the two independent spin variables and defining $\mathbf P_r=(P_2,P_3)$, the equations of motion in Cartesian components become:
\begin{subequations}\label{eq:massive-eom}
	\begin{gather}
		\sqrt G\,\dot{\mathbf x}
		=\frac{\mathbf P}{E_P}
		+\frac{e}{2mE_P}
		\left(S_0\mathbf B+\mathbf E\times\mathbf S\right),
		\label{eq:massive-x}
		\\
		\sqrt G\,\dot{\mathbf P}
		=e\mathbf E
		+\frac{e}{E_P}\mathbf P\times\mathbf B
		+\frac{e^2}{2mE_P}(\mathbf E\cdot\mathbf B)\mathbf S,
		\label{eq:massive-p}
		\\
		\begin{aligned}
			\sqrt G\,\dot{\mathbf S}_{r}
			=-\frac{e\mathbf P_{r}}{m^2E_P}
			\Big[
			&S_0\,\mathbf E\cdot\mathbf P
			-E_P\,\mathbf E\cdot\mathbf S\\
			&+\mathbf B\cdot(\mathbf P\times\mathbf S)
			-\frac{2es^2}{m}\mathbf E\cdot\mathbf B
			\Big].
		\end{aligned}
		\label{eq:massive-spin}
	\end{gather}
\end{subequations}
\noindent where $E_P=P_0=\sqrt{\mathbf P^2+m^2}$ and $\sqrt G=1+e\,\mathbf B\cdot\mathbf S/(2mE_P)$. The remaining components $S_0$ and $S_1$ are determined locally, on a chosen branch, by
	\begin{equation}
		P_0S_0=\mathbf P\cdot\mathbf S,
		\qquad
		(S_0)^2-|\mathbf S|^2=-4s^2.
		\label{eq:spin-constraints}
\end{equation}
The particle velocity is generally nonparallel to its momentum, analogous to the anomalous velocity in chiral kinetic theory. This arises from the interplay between the spin constraints and the external electromagnetic field. The spin also evolves without a magnetic dipole coupling, reflecting Thomas precession, but with corrections associated with the modified velocity momentum relation and the factor $\sqrt G$.
The collisionless kinetic equation for $f(t,\mathbf x,\mathbf P,\mathbf S_r)$ then reads
\begin{equation}
	\frac{\dif f}{\dif t}
		=\left(
		\partial_t
		+\dot{\mathbf x}\cdot\nabla_{\mathbf x}
		+\dot{\mathbf P}\cdot\nabla_{\mathbf P}
		+\dot{\mathbf S}_r\cdot\nabla_{\mathbf S_r}
		\right)f=0.
	\label{eq:reduced-kinetic}
\end{equation}
\noindent where $(\dot{\vecs{x}},\dot{\vecs{P}},\dot{\vecs{S}_r})$ are given by Eqs.~\eqref{eq:massive-x}--\eqref{eq:massive-spin}.

To determine the generalized Poisson structure and invariant measure on the reduced phase space, the gauge conditions $x^0-\lambda=0$ and $\omega^0=0$ are imposed together with the six original constraints, forming a complete second class set. The resulting full Dirac brackets define the generalized Poisson brackets below, which are again independent of the auxiliary spin gauge choice provided the combined constraint set remains second class.  Defining $\Omega_i=\frac{S_i}{2mE_P}$,
	the brackets take the form
\begin{subequations}\label{eq:massive-brackets}
		\begin{gather}
			\GPB{x_i,x_j}=\frac{\epsilon_{ijk}\Omega_k}{\sqrt G},\\
			\GPB{P_i,P_j}=-\frac{e\epsilon_{ijk}B_k}{\sqrt G},\\
			\GPB{P_i,x_j}=\frac{\delta_{ij}+e\Omega_iB_j}{\sqrt G},\\
			\GPB{x_i,S_j}=-\frac{P_j}{m^2\sqrt G}
			\bigl(
			S_i-\tfrac{S_0P_i}{E_P}
			+\tfrac{2es^2}{mE_P}B_i
			\bigr),\displaybreak[4]\\
			\GPB{P_i,S_j}=-\frac{eP_j}{m^2\sqrt G}
			\epsilon_{ikl}
			\left(S_k-\frac{S_0P_k}{E_P}\right)B_l,\\
			\GPB{S_i,S_j}=\frac{2}{m}\epsilon_{ijk}
			\left(E_PS_k-P_kS_0\right).
		\end{gather}
\end{subequations}
Here $S_\mu S^\mu=-4s^2$ has been used. Only $S_2$ and $S_3$ are independent, with $S_0$ and $S_1$ determined by the constraints in~\eqref{eq:spin-constraints}.

For the independent physical variables $\zeta_r^a=(\mathbf x,\mathbf P,\mathbf S_{\text{r}})$, these GPBs define the Poisson structure $\Pi^{ab}=\{\zeta_r^a,\zeta_r^b\}_{\mathrm{GPB}}$. Obtained from the full Dirac brackets, they satisfy the Jacobi identity \cite{Diramanifold}. Wherever $\Pi^{ab}$ is nondegenerate, its inverse $\Omega_{ab}=(\Pi^{-1})_{ab}$ defines the symplectic two-form $\Omega=\frac12\Omega_{ab}\, \dif \zeta_r^a\wedge \dif \zeta_r^b$. The associated symplectic volume gives the invariant phase space measure,
\begin{equation}
		\begin{aligned}
			\dif\Gamma
			&=\rho\,\dif^3\mathbf x\,\dif ^3\mathbf P\,\dif^2\mathbf S_{\text{r}},\\
			\rho
			&=\sqrt{|\det\Pi^{-1}|}
			=\frac{m|\sqrt G|}{2|P_0S_1-P_1S_0|}.
		\end{aligned}
		\label{eq:measure-density}
\end{equation}
Here $\rho$ is the phase space measure density.
Introducing the Lorentz scalar spin measure, with the same covariant form as in Ref. \cite{Weickgenannt:2020aaf} but our normalization,
\begin{equation}
		\begin{aligned}
			\dif S(P)
			&=m\,\dif^4S\,
			\delta(S_\mu S^\mu+4s^2)\,
			\delta(P_\mu S^\mu)\\
			&=\dif S_2\,\dif S_3\frac{m}
			{2|P_0S_1-P_1S_0|},
		\end{aligned}
		\label{eq:spin-measure}
\end{equation}
the invariant phase space measure becomes
\begin{equation}
		\dif\Gamma
		=\dif^3\mathbf x\,\dif^3\mathbf P\,\dif S(P)\,|\sqrt G|.
		\label{eq:invariant-measure}
\end{equation}

The same measure follows from the unified prescription \cite{Garczynski:1986fw}, combining Faddeev's gauge-fixed measure for first class constraints \cite{Faddeev:1969su} with Senjanovic's measure for second class constraints \cite{Senjanovic},
\begin{equation}
    \begin{aligned}
        \dif\mu_{\mathrm{FS}}
        ={}&\dif^4x\,\dif^4P\,\dif^4\omega\,\dif^4\pi\\
        &\times\prod_i\delta(\xi_i)\,
        \sqrt{\left|\det\!\left(\{\xi_i,\xi_j\}_{\mathrm{PB}}\right)\right|},
    \end{aligned}
    \label{eq:lcm}
\end{equation}
where $\xi_i$ comprise the first and second class constraints together with the gauge conditions. Performing a change of variables and the delta-function integrations reproduces  Eq.~\eqref{eq:invariant-measure}, providing an equivalent representation in the extended phase space.

A consistent kinetic description requires this measure to be preserved during particle evolution. For a time-independent gauge potential, the symplectic structure of the reduced phase space guarantees preservation of the measure under Hamiltonian flow. For a time-dependent gauge potential, the explicit time dependence of $\rho$ must be included. Substituting the equations of motion and using the homogeneous Maxwell equations, a direct calculation yields the Liouville identity (see Supplemental Material, Sec. \ref{sm:cl}),
	\begin{equation}
		\partial_t\rho+\partial_{\zeta_r^a}\!\left(\rho\dot\zeta_r^a\right)=0.
		\label{eq:liouville}
\end{equation}
This identity holds exactly within the present formulation and, together with the collisionless kinetic equation, ensures particle number conservation.

\prlheading{Conservation laws and balance equations}
The particle equations of motion and the preserved phase space measure provide a consistent basis for defining macroscopic currents. Writing $y^I=(\mathbf P,\mathbf S_r)$ and $\dif^5y=\dif^3\mathbf P\,\dif^2\mathbf S_r$, the charge current, energy-momentum tensor, total angular momentum current, and Boltzmann entropy current are defined by
\begin{equation}
	\begin{aligned}
		&\{J^\mu,T^{\mu\nu},J^{\mu\alpha\beta},E^\mu\}
		\\
		&=\int \dif^5y\,\rho\,\dot x^\mu f\,
		\{e,P^\nu,\mathcal J^{\alpha\beta},1-\ln f\}.
	\end{aligned}
	\label{eq:currents}
\end{equation}
\noindent where $\dot x^\mu=(1,\dot{\mathbf x})$ and $\mathcal J^{\alpha\beta}=x^{[\alpha}P^{\beta]}+S^{\alpha\beta}/2$ is the single-particle total angular momentum.

For a phase space function $A$, introduce the moment current $C_A^\mu=\int \dif^5y\,\rho\,\dot x^\mu Af$. The Liouville identity and the collisionless kinetic equation imply (see Supplemental Material, Sec. \ref{sm:cl})
\begin{equation}
	\begin{aligned}
		\partial_\mu C_A^\mu
		&=\int \dif^5y\,\rho f\,\frac{\dif A}{\dif t},\\
		\frac{\dif A}{\dif t}
		&=\partial_tA+\dot\zeta_r^a\partial_aA.
	\end{aligned}
	\label{eq:moment-identity}
\end{equation}

Since $\dif f/\dif t=0$, choosing $A=e$ and $A=1-\ln f$ gives charge and entropy conservation,
\begin{equation}
	\partial_\mu J^\mu=0,
	\qquad
	\partial_\mu E^\mu=0.
	\label{eq:charge-entropy}
\end{equation}
The equations of motion in Eqs.~\eqref{eq:massive-cov-x}--\eqref{eq:massive-cov-spin}, expressed in coordinate time, imply
\begin{equation}
		\begin{aligned}
			\frac{\dif P^\nu}{\dif t}
			&=eF^{\nu\lambda}\dot x_\lambda,\\
			\frac{\dif \mathcal J^{\alpha\beta}}{\dif t}
			&=e\,x^{[\alpha}F^{\beta]\lambda}\dot x_\lambda.
		\end{aligned}
		\label{eq:particle-balances}
\end{equation}
Substitution into Eq.~\eqref{eq:moment-identity} yields the energy-momentum and angular momentum balance equations,
\begin{equation}
	\begin{aligned}
		\partial_\mu T^{\mu\nu}
		&=F^{\nu\lambda}J_\lambda,\\
		\partial_\mu J^{\mu\alpha\beta}
		&=x^{[\alpha}F^{\beta]\lambda}J_\lambda.
	\end{aligned}
	\label{eq:macroscopic-balances}
\end{equation}
The right-hand sides represent the force and torque densities exerted by the external electromagnetic field on the matter.
Defining the spin angular momentum current as
$S^{\lambda\mu\nu}=\frac12\int \dif^5y\,\rho\,\dot x^\lambda S^{\mu\nu}f$,
Eqs.~\eqref{eq:currents} and~\eqref{eq:macroscopic-balances} imply
\begin{equation}
   T^{[\mu\nu]}+\partial_\lambda S^{\lambda\mu\nu}=0,
\end{equation}
which describes the local conversion between spin and orbital angular momentum.

\prlheading{Global equilibrium}
To examine global equilibrium within this kinetic theory, consider the local equilibrium ansatz \cite{Becattini_2013,PhysRevC.97.041901,FLORKOWSKI2019103709},
\begin{equation}
    f_{\mathrm{LE}}
    =f\Bigl[
    a(x)+b^\mu(x)P_\mu
    +\tfrac12\omega_{\mu\nu}(x)\mathcal J^{\mu\nu}
    \Bigr],
    \label{eq:equilibrium-ansatz}
\end{equation}
\noindent where the functional form of $f$ specifies the statistics. The coefficients depend only on spacetime coordinates, and $\omega_{\mu\nu}=-\omega_{\nu\mu}$. Substitution into the covariant collisionless kinetic equation~\eqref{eq:covariant-kinetic} gives the following sufficient conditions for $\dif f_{\mathrm{LE}}/\dif \lambda=0$ (see Supplemental Material, Sec. \ref{sm:eql}):
\begin{subequations}\label{eq:equilibrium-conditions}
	\begin{gather}
		\partial_\mu a-eF_{\mu\nu}\beta^\nu=0,
		\label{eq:equilibrium-a}
		\\
		\partial_{(\mu}\beta_{\nu)}=0,
		\label{eq:equilibrium-killing}
		\\
		\partial_\mu\omega_{\rho\sigma}=0,
		\label{eq:equilibrium-spin}
		\\
		F_{[\nu(\rho}\partial_{\mu]}\beta_{\alpha)}
		-F_{[\nu(\rho}\omega_{\mu]\alpha)}=0,
		\label{eq:equilibrium-field}
	\end{gather}
\end{subequations}
\noindent where $\beta^\mu=b^\mu+\omega^{\nu\mu}x_\nu$ is the thermal four vector. Equation~\eqref{eq:equilibrium-killing} is the Killing condition, while Eq.~\eqref{eq:equilibrium-a} relates the chemical potential gradient to the electromagnetic field.

If the electromagnetic field tensors at two spacetime points are both nonzero and are not proportional to each other, Eqs.~\eqref{eq:equilibrium-killing}--\eqref{eq:equilibrium-field} further imply (see Supplemental Material, Sec. \ref{sm:eql})
\begin{equation}
	\omega_{\mu\nu}
	=\frac12\partial_{[\mu}\beta_{\nu]}.
	\label{eq:spin-thermal-vorticity}
\end{equation}
Under these field conditions, the spin polarization potential equals the thermal vorticity, and the resulting global equilibrium conditions agree with those obtained from covariant statistical mechanics \cite{Becattini:2012StatMech,Florkowski:2018} and the Wigner function approach \cite{Weickgenannt:2019dks,Liu_2020}.

\prlheading{Conclusion}
In this Letter, we develop a Hamiltonian formulation of collisionless kinetic theory for relativistic spin-1/2 fermions in external electromagnetic fields. The formulation yields both a covariant kinetic equation and a kinetic equation for the independent physical variables in the reduced phase space. For chiral particle, it reproduces the standard chiral kinetic equations, noncanonical Poisson brackets, and invariant measure. For massive spin-1/2 fermions, the interplay between spin constraints and the electromagnetic field makes the velocity generally nonparallel to the kinetic momentum and modifies momentum and spin evolution. The reduced phase space exhibits noncanonical Poisson brackets and a modified invariant measure, analogous to those in chiral kinetic theory. 

The resulting kinetic equation and Liouville identity ensure exact conservation of particle number and Boltzmann entropy within the present formulation.
The energy-momentum and total angular momentum balance equations identify the force and torque densities exerted by the external field. Under the stated electromagnetic field conditions, the global equilibrium conditions agree with those obtained from covariant statistical mechanics and the Wigner function approach.

This framework provides a basis for studying coupled position, momentum, and spin transport in electromagnetic backgrounds, with potential applications to condensed matter systems, plasmas, and heavy-ion collisions, particularly away from equilibrium. Magnetic dipole coupling and collision terms consistent with the phase space structure and conservation laws remain to be incorporated and are left for future work.

 The authors thank RenGuo Wang, XiaoYu Li, and Jai Prakash for helpful discussions. This work was supported by the National Natural Science Foundation of China (NSFC) under Grant Nos. 12422508, 12375124, and the Science and Technology Commission
of Shanghai Municipality under Grant No. 23JC1402700. Y.S. thanks the sponsorship from Yangyang Development Fund.

\bibliography{biblio_paper}{}

\clearpage
\onecolumngrid
\begin{center}
    \normalfont\normalsize\bfseries
    SUPPLEMENTAL MATERIAL FOR ``CONSTRAINED HAMILTONIAN FORMULATION OF KINETIC THEORY FOR RELATIVISTIC SPIN-$1/2$ PARTICLES''\par
\end{center}
\medskip

\setcounter{equation}{0}
\setcounter{figure}{0}
\setcounter{table}{0}
\setcounter{page}{1}
\setcounter{section}{0}
\renewcommand{\thesection}{\Alph{section}}
\renewcommand{\theequation}{S\arabic{equation}}
\renewcommand{\thefigure}{S\arabic{figure}}
\makeatletter
\renewcommand{\bibnumfmt}[1]{[S#1]}
\def\section{\@startsection{section}{1}{\z@}
    {3ex \@plus 1ex \@minus .2ex}{1.5ex}
    {\normalfont\normalsize\bfseries\centering}}
\def\@hangfrom@section#1#2#3{\@hangfrom{#1#2}#3}
\makeatother

\section{Massless spin-1/2 fermion: Dirac--Bergmann analysis and generalized Poisson brackets}
\label{SM:ckt}
In the constrained Hamiltonian formalism, the evolution of any phase space function $F$ without explicit dependence on $\lambda$ is governed by
\begin{equation}
    \frac{\dif F}{\dif\lambda}=\PB{F,H_p},
\end{equation}
where $H_p$ denotes the primary Hamiltonian. In the present formulation with worldline reparametrization invariance, the canonical Hamiltonian vanishes, so $H_p$ is a linear combination of the primary constraints. For massless spin-$1/2$ particles with the constraints in Eq.~\eqref{eq:massless-constraints}, it takes the form
\begin{equation}
    H_p=\frac{\lambda_1}{2}P^2
    +\frac{\lambda_2}{2}(\pi^2-a_2)
    +\frac{\lambda_3}{2}(\omega^2-a_3)
    +\lambda_4P\cdot\omega
    +\lambda_5P\cdot\pi
    +\lambda_6\omega\cdot\pi
    +\lambda_7N\cdot\omega
    +\lambda_8N\cdot\pi.
\end{equation}

The Dirac--Bergmann algorithm requires the constraints to be preserved under evolution \cite{Dirac:1958sq,Bergmann:1954tc,Wipf,Brown:2022gha},
\begin{equation}
    \frac{\dif T_i}{\dif\lambda}
    =\PB{T_i,H_p}\approx0,
\end{equation}
where $\approx$ denotes weak equality, namely equality on the constrained phase space. These conditions determine Lagrange multipliers or generate further constraints, and are imposed iteratively until no additional constraints arise.

Introducing
\begin{equation}
    c_1=F_{\mu\nu}P^\mu\omega^\nu,\quad
    c_2=F_{\mu\nu}P^\mu\pi^\nu,\quad
    c_3=F_{\mu\nu}\omega^\mu\pi^\nu,\quad
    d_1=N\cdot P,\quad d_2=N^2,
\end{equation}
the consistency conditions for the eight primary constraints become
\begin{equation}
    \begin{aligned}
        \frac{\dif}{\dif\lambda}\{T_1,\ldots,T_8\}
        &\approx\Bigl\{
        e(\lambda_4c_1+\lambda_5c_2),\,
        -\lambda_6a_2,\,
        \lambda_6a_3,\,
        -e\lambda_1c_1+e\lambda_5c_3+\lambda_8d_1,\\
        &\qquad
        -e\lambda_1c_2-e\lambda_4c_3-\lambda_7d_1,\,
        \lambda_2a_2-\lambda_3a_3,\,
        \lambda_5d_1+\lambda_8d_2,\,
        -\lambda_4d_1-\lambda_7d_2
        \Bigr\}=\mathbf 0.
    \end{aligned}
\end{equation}
These conditions determine six multipliers in terms of the arbitrary functions $\lambda_1$ and $\lambda_2$, without generating additional constraints. The eight primary constraints therefore form the complete set.

To classify these constraints, define their Poisson bracket matrix
$C_{ij}=\PB{T_i,T_j}$. On the constrained phase space, this matrix is
\begin{equation}
    C_{ij}\approx
    \begin{pmatrix}
        0 & 0 & 0 & ec_1 & ec_2 & 0 & 0 & 0\\
        0 & 0 & 0 & 0 & 0 & -a_2 & 0 & 0\\
        0 & 0 & 0 & 0 & 0 & a_3 & 0 & 0\\
        -ec_1 & 0 & 0 & 0 & ec_3 & 0 & 0 & d_1\\
        -ec_2 & 0 & 0 & -ec_3 & 0 & 0 & -d_1 & 0\\
        0 & a_2 & -a_3 & 0 & 0 & 0 & 0 & 0\\
        0 & 0 & 0 & 0 & d_1 & 0 & 0 & d_2\\
        0 & 0 & 0 & -d_1 & 0 & 0 & -d_2 & 0
    \end{pmatrix}.
\end{equation}
In the regular region where $\Delta\neq0$, this matrix has rank six. The eight constraints can therefore be organized into six second class and two first class combinations.

A convenient choice of second class constraints is
\begin{equation}
    \{\xi_1,\ldots,\xi_6\}
    =\left\{
    \frac12(\omega^2-a_3),\,
    P\cdot\omega,\,
    P\cdot\pi,\,
    \omega\cdot\pi,\,
    N\cdot\omega,\,
    N\cdot\pi
    \right\},
\end{equation}
whose Poisson bracket matrix is nondegenerate. After imposing these constraints strongly and redefining the remaining multipliers, the partially reduced Hamiltonian takes the form given in Eq.~\eqref{eq:massless-hamiltonian}, with subsequent evolution generated by the corresponding Dirac bracket.

For any two phase space functions $F$ and $G$, the Dirac bracket is defined by \cite{Dirac:1958sq}
\begin{equation}
    \DB{F,G}
    =\PB{F,G}
    -\PB{F,\xi_i}
    (\mathcal C^{-1})_{ij}
    \PB{\xi_j,G},
\end{equation}
where $\mathcal C_{ij}=\PB{\xi_i,\xi_j}$ and
$i,j=1,\ldots,6$. 
On the constrained phase space, the Dirac brackets among
$(x^\mu,P^\mu,\omega^\mu,\pi^\mu)$ take the form
\begin{subequations}
    \label{eq:DBr}
    \begin{gather}
        \DB{x^\alpha,x^\beta}
        =\frac{d_2(\omega^\alpha\pi^\beta-\pi^\alpha\omega^\beta)}
        {\Delta},
        \qquad
        \DB{x^\alpha,P^\beta}
        =g^{\alpha\beta}
        +\frac{ed_2F^{\beta\mu}
        (\omega^\alpha\pi_\mu-\pi^\alpha\omega_\mu)}
        {\Delta},
        \\
        \DB{x^\alpha,\omega^\beta}
        =\frac{\omega^\alpha(d_2P^\beta-d_1N^\beta)}{\Delta},
        \qquad
        \DB{x^\alpha,\pi^\beta}
        =\frac{\pi^\alpha(d_2P^\beta-d_1N^\beta)}{\Delta},
        \\[3pt]
        \DB{P^\alpha,P^\beta}
        =eF^{\alpha\beta}
        +\frac{e^2d_2F^{\alpha\rho}F^{\beta\sigma}
        (\omega_\rho\pi_\sigma-\pi_\rho\omega_\sigma)}
        {\Delta},
        \\
        \DB{P^\alpha,\omega^\beta}
        =\frac{eF^{\alpha\mu}\omega_\mu
        (d_2P^\beta-d_1N^\beta)}{\Delta},
        \qquad
        \DB{P^\alpha,\pi^\beta}
        =\frac{eF^{\alpha\mu}\pi_\mu
        (d_2P^\beta-d_1N^\beta)}{\Delta},
        \\[3pt]
        \DB{\omega^\alpha,\omega^\beta}=0,
        \qquad
        \DB{\omega^\alpha,\pi^\beta}
        =g^{\alpha\beta}
        +\frac{
        d_2P^\alpha P^\beta
        -d_1(P^\alpha N^\beta+N^\alpha P^\beta)
        +ec_3N^\alpha N^\beta
        }{\Delta}
        -\frac{\omega^\alpha\omega^\beta}{a_3},
        \\[3pt]
        \DB{\pi^\alpha,\pi^\beta}
        =\frac{\omega^\beta\pi^\alpha-\omega^\alpha\pi^\beta}{a_3}.
    \end{gather}
\end{subequations}
All remaining brackets follow by antisymmetry. Together with $\mathcal H_{\mathrm{pr}}$, these brackets reproduce the covariant equations of motion given in Eqs.~\eqref{eq:massless-covariant}.

To obtain the generalized Poisson brackets and invariant measure in terms of the six independent physical phase space variables $(\mathbf x,\mathbf P)$, the remaining gauge freedoms must be fixed. A covariant gauge choice keeps Lorentz covariance manifest. However, an admissible gauge fixing only selects a representative of each gauge orbit and leaves the underlying covariant dynamics unchanged. The gauge conditions can therefore be chosen as $x^0-\lambda=0$ and $\omega^1=0$, with $\lambda=t$. Together with the eight constraints in Eq.~\eqref{eq:massless-constraints}, these conditions form the complete second class set
\begin{equation}
    \{\xi_1^{\mathrm{full}},\ldots,\xi_{10}^{\mathrm{full}}\}
    =\{T_1,\ldots,T_8,\;x^0-\lambda,\;\omega^1\}.
\end{equation}
Defining the full bracket matrix
\begin{equation}
    \mathcal C_{\mathrm{full},IJ}
    =\PB{\xi_I^{\mathrm{full}},\xi_J^{\mathrm{full}}},
    \qquad I,J=1,\ldots,10,
\end{equation}
the full Dirac bracket is
\begin{equation}
    \{F,G\}_{\mathrm{DB,full}}
    =\PB{F,G}
    -\PB{F,\xi_I^{\mathrm{full}}}
    (\mathcal C_{\mathrm{full}}^{-1})_{IJ}
    \PB{\xi_J^{\mathrm{full}},G}.
\end{equation}
The generalized Poisson brackets on the reduced phase space are defined by $\{F,G\}_{\mathrm{GPB}}=\{F,G\}_{\mathrm{DB,full}}$.

With $N^\mu=(1,0,0,0)$ and spatial components expressed in Cartesian coordinates ($x_i\equiv x^i$ and $P_i\equiv P^i$), the generalized Poisson brackets among $(\mathbf x,\mathbf P)$ take the form
\begin{equation}
    \GPB{x_i,x_j}=\frac{\epsilon_{ijk}\Omega_k}{\sqrt G},\qquad
    \GPB{P_i,P_j}=-\frac{e\epsilon_{ijk}B_k}{\sqrt G},\qquad
    \GPB{P_i,x_j}=\frac{\delta_{ij}+e\Omega_iB_j}{\sqrt G},
    \label{eq:massless-full-brackets}
\end{equation}
where $\boldsymbol\Omega$ and $\sqrt G$ are defined in Eq.~\eqref{eq:ckt-definitions}. The GPB matrix $\Pi$ of $(\mathbf x,\mathbf P)$ determines the invariant measure through
\begin{equation}
    \dif\Gamma
    =\sqrt{|\det\Pi^{-1}|}\,\dif^3\mathbf x\,\dif^3\mathbf P.
\end{equation}
Substituting Eq.~\eqref{eq:massless-full-brackets} gives $\sqrt{|\det\Pi^{-1}|}=|\sqrt G|$, reproducing the measure quoted in the main text.

\section{Massive spin-$1/2$ fermions: Dirac--Bergmann analysis and generalized Poisson brackets}
\label{sm:skm}
For massive spin-$1/2$ particles, the six constraints in Eq.~\eqref{eq:massive-constraints} give the primary Hamiltonian
\begin{equation}
    H_p=\frac{\lambda_1}{2}(P^2-m^2)
    +\frac{\lambda_2}{2}(\pi^2-a_2)
    +\frac{\lambda_3}{2}(\omega^2-a_3)
    +\lambda_4P\cdot\omega
    +\lambda_5P\cdot\pi
    +\lambda_6\omega\cdot\pi.
\end{equation}
Following the consistency procedure in Sec.~\ref{SM:ckt} and using the definitions of $c_1$, $c_2$, and $c_3$ given there, the preservation conditions become
\begin{equation}
    \begin{aligned}
        \frac{\dif}{\dif\lambda}\{T_1,\ldots,T_6\}
        &\approx\Bigl\{
        e(\lambda_4c_1+\lambda_5c_2),\,
        -\lambda_6a_2,\,
        \lambda_6a_3,\,
        -e\lambda_1c_1+\lambda_5\Delta_1,\,
        -e\lambda_1c_2-\lambda_4\Delta_1,\,
        \lambda_2a_2-\lambda_3a_3
        \Bigr\}=\mathbf 0,
    \end{aligned}
\end{equation}
where $\Delta_1=m^2+ec_3=m^2+eF_{\mu\nu}S^{\mu\nu}/4$, as defined in the main text. For $\Delta_1\neq0$, the consistency conditions generate no additional constraints, so the six primary constraints form the complete set. 

On the constrained phase space, their Poisson bracket matrix has rank four and takes the form
\begin{equation}
    C_{ij}\approx
    \begin{pmatrix}
        0 & 0 & 0 & ec_1 & ec_2 & 0\\
        0 & 0 & 0 & 0 & 0 & -a_2\\
        0 & 0 & 0 & 0 & 0 & a_3\\
        -ec_1 & 0 & 0 & 0 & \Delta_1 & 0\\
        -ec_2 & 0 & 0 & -\Delta_1 & 0 & 0\\
        0 & a_2 & -a_3 & 0 & 0 & 0
    \end{pmatrix},
\end{equation}
so the six constraints can be organized into four second class and two first class combinations. 

The second class constraints can be chosen as
\begin{equation}
    \{\xi_1,\ldots,\xi_4\}
    =\left\{
    \frac12(\omega^2-a_3),\,
    P\cdot\omega,\,
    P\cdot\pi,\,
    \omega\cdot\pi
    \right\},
\end{equation}
whose Poisson bracket matrix is nondegenerate. Imposing these constraints strongly and redefining the remaining multipliers yields $\mathcal H_{\mathrm{pr}}$ in Eq.~\eqref{eq:massive-hamiltonian}.

The Dirac bracket is constructed as in Sec.~\ref{SM:ckt}, now with $i,j=1,\ldots,4$. The brackets among $(x^\mu,P^\mu,\omega^\mu,\pi^\mu)$ on the constrained phase space take the form
\begin{subequations}
    \label{eq:DBrm}
    \begin{gather}
        \DB{x^\mu,x^\nu}
        =\frac{\pi^\mu\omega^\nu-\omega^\mu\pi^\nu}{\Delta_1},
        \qquad
        \DB{x^\mu,P^\nu}
        =g^{\mu\nu}
        +\frac{eF^{\nu\rho}
        (\pi^\mu\omega_\rho-\omega^\mu\pi_\rho)}{\Delta_1},
        \\
        \DB{x^\mu,\omega^\nu}
        =-\frac{\omega^\mu P^\nu}{\Delta_1},
        \qquad
        \DB{x^\mu,\pi^\nu}
        =-\frac{\pi^\mu P^\nu}{\Delta_1},
        \\
        \DB{P^\mu,P^\nu}
        =eF^{\mu\nu}
        +\frac{e^2F^{\mu\rho}F^{\nu\sigma}
        (\pi_\rho\omega_\sigma-\omega_\rho\pi_\sigma)}
        {\Delta_1},
        \\
        \DB{P^\mu,\omega^\nu}
        =-\frac{eF^{\mu\rho}\omega_\rho P^\nu}{\Delta_1},
        \qquad
        \DB{P^\mu,\pi^\nu}
        =-\frac{eF^{\mu\rho}\pi_\rho P^\nu}{\Delta_1},
        \\
        \DB{\omega^\mu,\omega^\nu}=0,
        \qquad
        \DB{\omega^\mu,\pi^\nu}
        =g^{\mu\nu}
        -\frac{P^\mu P^\nu}{\Delta_1}
        -\frac{\omega^\mu\omega^\nu}{a_3},
        \qquad
        \DB{\pi^\mu,\pi^\nu}
        =\frac{\pi^\mu\omega^\nu-\omega^\mu\pi^\nu}{a_3}.
    \end{gather}
\end{subequations}
Together with $\mathcal H_{\mathrm{pr}}$, these brackets give the auxiliary spin evolution
\begin{equation}
    \frac{\dif\omega^\mu}{\dif\lambda}
    =\frac{e\lambda_1}{\Delta_1}
    F_{\alpha\beta}P^\alpha\omega^\beta P^\mu
    +\lambda_2\pi^\mu,
    \qquad
    \frac{\dif\pi^\mu}{\dif\lambda}
    =\frac{e\lambda_1}{\Delta_1}
    F_{\alpha\beta}P^\alpha\pi^\beta P^\mu
    -\frac{a_2}{a_3}\lambda_2\omega^\mu.
\end{equation}
The $\lambda_2$ terms cancel in the evolution of both the physical spin tensor $S^{\mu\nu}=2\omega^{[\mu}\pi^{\nu]}$ and the spin four-vector $S^\mu$ defined in Eq.~\eqref{eq:spin-vector}. The resulting covariant equations reproduce Eqs.~\eqref{eq:massive-covariant} and are independent of the gauge freedom of the auxiliary spin variables.

To obtain the Poisson structure in terms of the eight independent physical phase space variables, the remaining gauge freedoms are fixed by $x^0-\lambda=0$ and $\omega^0=0$, with $\lambda=t$. The complete second class set is
\begin{equation}
    \{\xi_1^{\mathrm{full}},\ldots,\xi_8^{\mathrm{full}}\}
    =\{T_1,\ldots,T_6,\;x^0-\lambda,\;\omega^0\}.
\end{equation}
The full Dirac bracket is constructed from this set as in Appendix~\ref{SM:ckt}; on the physical reduced variables, it defines the GPB. 

Using the Cartesian component convention of the main text, the full Dirac brackets reproduce the generalized Poisson brackets in Eqs.~\eqref{eq:massive-brackets}. For the GPB matrix $\Pi$ of $(\mathbf x,\mathbf P,\mathbf S_r)$, direct calculation gives
\begin{equation}
    \det\Pi^{-1}
    =\frac{m^2G}{4(E_PS_1-P_1S_0)^2}.
\end{equation}
Taking the square root of its absolute value reproduces the measure density in Eq.~\eqref{eq:measure-density}.

\section{Liouville identity, conservation laws, and balance equations}
\label{sm:cl}

To verify the Liouville identity, solve the spin
constraints~\eqref{eq:spin-constraints} for $S_0$ and $S_1$:
\begin{equation}
    S_0=\frac{E_P(P_2S_2+P_3S_3)\pm P_1\sqrt R}
    {m^2+P_2^2+P_3^2},
    \qquad
    S_1=\frac{P_1(P_2S_2+P_3S_3)\pm E_P\sqrt R}
    {m^2+P_2^2+P_3^2},
\end{equation}
where
\begin{equation}
    E_P=\sqrt{\mathbf P^2+m^2},\qquad
    R=(P_2S_2+P_3S_3)^2
    +(m^2+P_2^2+P_3^2)(4s^2-S_2^2-S_3^2).
\end{equation}
The signs are chosen consistently on either branch with $R>0$.

The measure density defined in the main text can be written as
$\rho=Q|\sqrt G|$, where $Q=m/(2|E_PS_1-P_1S_0|)$ is a function
of the independent variables $(\mathbf P,S_2,S_3)$. All derivatives below account for the
dependence of $E_P$, $S_0$, and $S_1$ on these independent
variables. Wherever $\sqrt G$ is nonzero, its sign is locally
fixed, so it suffices to prove the Liouville identity for
$Q\sqrt G$.

Using Eq.~\eqref{eq:massive-x}, the time and spatial contributions give
\begin{equation}
    \partial_t(Q\sqrt G)
    +\nabla_{\mathbf x}\cdot(Q\sqrt G\,\dot{\mathbf x})
    =\frac{eQ}{2mE_P}\Bigl[
    \mathbf S\cdot(\partial_t\mathbf B+\nabla_{\mathbf x}\times\mathbf E)
    +S_0\nabla_{\mathbf x}\cdot\mathbf B
    \Bigr]=0,
\end{equation}
by the homogeneous Maxwell equations.
For the momentum and spin contributions, substituting the
expressions above into Eqs.~\eqref{eq:massive-p}
and~\eqref{eq:massive-spin} and differentiating gives
\begin{equation}
    \nabla_{\mathbf P}\cdot
    \left(Q\sqrt G\,\dot{\mathbf P}\right)
    +\sum_{r=2}^{3}\frac{\partial}{\partial S_r}
    \left(Q\sqrt G\,\dot S_r\right)=0.
\end{equation}
This cancellation can be verified symbolically in Mathematica. Combining these results establishes the Liouville identity~\eqref{eq:liouville}.

Combining the Liouville identity with the collisionless
kinetic equation~\eqref{eq:reduced-kinetic} gives the
phase space continuity equation
\begin{equation}
    \frac{\partial(\rho f)}{\partial t}
    +\frac{\partial}{\partial\zeta_r^a}
    \left(\rho\dot\zeta_r^a f\right)=0.
\end{equation}
Multiplying this equation by a phase space function
$A(t,\zeta_r)$ and applying the product rule yields
\begin{equation}
    \frac{\partial(\rho Af)}{\partial t}
    +\frac{\partial}{\partial\zeta_r^a}
    \left(\rho\dot\zeta_r^a Af\right)
    =\rho f\,\frac{\dif A}{\dif t},
    \qquad
    \frac{\dif A}{\dif t}
    =\partial_t A+\dot\zeta_r^a\partial_a A.
\end{equation}
Integrating over $\dif^5y=\dif^3\mathbf P\,\dif^2\mathbf S_r$
at fixed $t$ and $\mathbf x$, and assuming zero net boundary
flux in momentum and spin, gives the local moment
equation~\eqref{eq:moment-identity}. The choices of $A$ in
the main text then yield the corresponding conservation laws
and balance equations.

\section{Global equilibrium}
\label{sm:eql}

For any choice of statistics, the distribution in
Eq.~\eqref{eq:equilibrium-ansatz} is constant along the
trajectory if its argument
$\Phi=a+b^\mu P_\mu+\tfrac12\omega_{\mu\nu}\mathcal J^{\mu\nu}$
satisfies $\dif\Phi/\dif\lambda=0$.
Using $\mathcal J^{\mu\nu}=x^{[\mu}P^{\nu]}+\tfrac12 S^{\mu\nu}$,
we combine the orbital term with $b^\mu P_\mu$ to obtain
\begin{equation}
    \Phi=a+\beta^\mu P_\mu+\frac14\omega_{\mu\nu}S^{\mu\nu},
    \qquad
    \beta^\mu=b^\mu+\omega^{\nu\mu}x_\nu.
    \label{eq:sm-equilibrium-argument}
\end{equation}
The particle equations of motion give
\begin{equation}
    \frac{\dif S^{\mu\nu}}{\dif\lambda}
    =2\left(
        P^\mu\frac{\dif x^\nu}{\dif\lambda}
        -P^\nu\frac{\dif x^\mu}{\dif\lambda}
    \right).
\end{equation}
Define
\begin{equation}
    d_{\mu\nu}=\partial_\mu\beta_\nu-\omega_{\mu\nu}.
    \label{eq:sm-equilibrium-difference}
\end{equation}
Taking the total derivative of $\Phi$ along the trajectory
and substituting the equations of motion for
$\dif P^\mu/\dif\lambda$ and $\dif S^{\mu\nu}/\dif\lambda$
gives
\begin{equation}
    \frac{\dif\Phi}{\dif\lambda}
    =\frac{\dif x^\mu}{\dif\lambda}
    \left[
        \bigl(\partial_\mu a-eF_{\mu\nu}\beta^\nu\bigr)
        +P^\nu d_{\mu\nu}
        +\frac14 S^{\rho\sigma}\partial_\mu\omega_{\rho\sigma}
    \right].
    \label{eq:sm-equilibrium-transport}
\end{equation}
Substituting $\dif x^\mu/\dif\lambda$ from
Eq.~\eqref{eq:massive-cov-x} gives
\begin{equation}
    \begin{aligned}
        \frac{\dif\Phi}{\dif\lambda}
        &=\lambda_1
        \left(P^\mu+\frac{e}{2\Delta_1}
        S^{\mu\nu}F_{\nu\rho}P^\rho\right)
        \bigl(\partial_\mu a-eF_{\mu\sigma}\beta^\sigma\bigr)
        +\lambda_1P^\mu P^\alpha d_{\mu\alpha}
        \\
        &\quad+\frac{e\lambda_1}{2\Delta_1}
        S^{\mu\nu}F_{\nu\rho}P^\rho P^\alpha d_{\mu\alpha}
        +\frac{\lambda_1}{4}
        \left(P^\mu+\frac{e}{2\Delta_1}
        S^{\mu\nu}F_{\nu\rho}P^\rho\right)
        S^{\alpha\beta}\partial_\mu\omega_{\alpha\beta}.
    \end{aligned}
\end{equation}
Equations~\eqref{eq:equilibrium-a} and~\eqref{eq:equilibrium-spin}
set the first and last lines to zero, respectively.
The Killing condition~\eqref{eq:equilibrium-killing} makes
$d_{\mu\alpha}$ antisymmetric, so
$P^\mu P^\alpha d_{\mu\alpha}=0$. The remaining term is
\begin{equation}
    \frac{\dif\Phi}{\dif\lambda}
    =\frac{e\lambda_1}{2\Delta_1}
    S^{\mu\nu}F_{\nu\rho}P^\rho P^\alpha d_{\mu\alpha}.
    \label{eq:sm-equilibrium-residual}
\end{equation}
The contraction with $S^{\mu\nu}P^\rho P^\alpha$ selects
the part of $F_{\nu\rho}d_{\mu\alpha}$ that is antisymmetric
in $\mu,\nu$ and symmetric in $\rho,\alpha$.
Equation~\eqref{eq:equilibrium-field} sets this part to zero,
giving $\dif\Phi/\dif\lambda=0$.

The Killing condition~\eqref{eq:equilibrium-killing} implies
$\partial_\lambda\partial_\mu\beta_\nu=0$, while
Eq.~\eqref{eq:equilibrium-spin} gives
$\partial_\lambda\omega_{\mu\nu}=0$. Thus $d_{\mu\nu}$ is
constant throughout the connected region under consideration.
For a nonzero electromagnetic field tensor, 
Eq.~\eqref{eq:equilibrium-field} implies
\begin{equation}
    d_{\mu\nu}=\kappa(x)F_{\mu\nu}(x),
    \label{eq:sm-equilibrium-proportionality}
\end{equation}
where $\kappa(x)$ is a scalar function. At the two points
$x_1$ and $x_2$ where the field tensors are nonzero, the
constancy of $d_{\mu\nu}$ gives
\begin{equation}
    d_{\mu\nu}
    =\kappa(x_1)F_{\mu\nu}(x_1)
    =\kappa(x_2)F_{\mu\nu}(x_2).
\end{equation}
When $F_{\mu\nu}(x_1)$ and $F_{\mu\nu}(x_2)$ are not
proportional, the above relation requires $d_{\mu\nu}=0$.
This gives $\omega_{\mu\nu}=\partial_\mu\beta_\nu$ and
reproduces Eq.~\eqref{eq:spin-thermal-vorticity} upon
using the Killing condition.

\end{document}